\documentclass{cjpsuppl}
 \usepackage{graphics,slashed} 
 \begin{document}
 \slacs{.6mm}
 \title{Antiproton polarization induced by lepton interactions}
 \authori{D.~S.~O'Brien$^*$\footnotetext{$^*$ email: donie@maths.tcd.ie} {\ } and{\ } N.~H.~Buttimore}
 \addressi{School of Mathematics, Trinity College, Dublin, Ireland}
 \authorii{}
 \addressii{}
 \authoriii{}   \addressiii{}
 \authoriv{}    \addressiv{}
 \authorv{}     \addressv{}
 \authorvi{}    \addressvi{}
 \headtitle{Antiproton polarization induced by lepton interactions}
 \headauthor{D.~S.~O'Brien, N.~H.~Buttimore}
 \lastevenhead{D.~S.~O'Brien, N.~H.~Buttimore: Antiproton polarization induced by lepton interactions}
 \refnum{}
 \daterec{} 
 \suppl{?}  \year{2007} \setcounter{page}{1}
 \maketitle

 \begin{abstract}
We present expressions for electromagnetic helicity amplitudes and spin observables for any elastic spin 1/2 - spin 1/2 scattering to first order in QED.  In particular all electromagnetic helicity amplitudes and spin observables for elastic antiproton-electron and antiproton-proton scattering via single $t$-channel photon exchange are presented.  Spin observables are required to describe the rate of increase of polarization in spin filtering.  The $\mathcal{PAX}$ collaboration at GSI Darmstadt is interested in the buildup of polarization of an antiproton beam by repeated interaction with a hydrogen gas target in a storage ring \cite{Rathmann:2004pm,Barone:2005pu}.  In order for the beam particles to remain in the ring after scattering it is important to consider  small angle scattering, hence small momentum transfer $t$.  In the low momentum transfer region electromagnetic effects dominate the hadronic effects.  Of immediate importance is whether the polarization of an antiproton beam can be built up by spin filtering off polarized electrons either in a target or in a beam.  We present the theoretical background for this discussion.
 \end{abstract}

\section{Introduction} 
\noindent
We consider electromagnetic helicity amplitudes and spin observables, because electromagnetic effects dominate the hadronic effects in the low momentum transfer region of interest where many of the beam particles remain in the storage ring after interaction with the target.  {\it Mathematica} and its add on package {\it Tracer} \cite{Jamin:1991dp} are used to directly compute helicity amplitudes and spin observables from the Feynman rules for single photon exchange in QED.  Using the most generic traces allows for any polarization phenomena to be analyzed by substituting in specific values for the spin four vectors $S_i$, where $i \,\in \,\{1,2,3,4\}$.  The general elastic process 
\begin{equation}
 A\,(\,p_1,\,S_1\,) + B\,(\,p_2,\,S_2\,) \longrightarrow A\,(\,p_3,\,S_3\,) + B\,(\,p_4,\,S_4\,)
\end{equation} 
is investigated, where it is assumed that the beam particles ($A$) of mass $M$ are antiprotons and the target particles ($B$) of mass $m$ are electrons or protons, the four momentum transfer being $q = p_3 - p_1 = p_2 - p_4$.  The target is initially polarized and the beam is initially unpolarized and we seek to model the buildup of polarization of the antiproton beam. 
\section{Definitions and normalization}
\noindent
The differential cross section for the elastic scattering of two spin 1/2 particles is related to the helicity amplitudes $\mathcal{M}\,(\Lambda'\,\lambda'\,;\Lambda\,\lambda\,)$
 by
\begin{equation}
s \,\frac{d\,\sigma}{d\,\Omega} =
\, \displaystyle{\frac{1}{(8\,\pi)^{\,2}}\sum_{\lambda\,\lambda'\,\Lambda\,\Lambda'} }\frac{1}{4}\,|\,\mathcal{M}\,(\,\Lambda'\,\lambda'\,;\,\Lambda\,\lambda\,)\,|^{\,2} 
\end{equation}
where $\lambda,\Lambda$ and $\lambda',\Lambda'$ are the helicities of the initial and final particles respectively.  The $s$, $t$ and $u$ in this paper are the Mandelstam variables.
\noindent
We  concentrate on elastic $2 \rightarrow 2$ scattering in the Centre-of-Mass frame where the momentum and spin four-vectors of the two initial and two scattered particles are:

\vspace*{-5ex}
\begin{table}
\begin{center}
\caption{\label{tab:5/tc} Momenta and spin 4-vectors in the Centre-of-Mass frame.}
\vspace{2mm}
\begin{tabular}{|ccc|ccc|} \hline
\multicolumn{6}{|c|}{} \\*[-2ex]
\multicolumn{6}{|c|}{Centre-of-Mass Momenta vectors} \\[0.5ex] \hline
& & & & &\\[-1ex]
$ p_1 $&$  = $&$ \left(\,E_1,\,0,\,0,\,k\,\right) $& $ p_3 $&$  = $&$ \left(\,E_1,\,k\,\sin\theta,\,0,\,k\,\cos\theta\,\right)$\\[1ex]
$ p_2 $&$  =$&$  \left(\,E_2,\,0,\,0,\,-k\,\right) $&$ p_4 $&$  = $&$ \left(\,E_2,\,-k\,\sin\theta,\,0,\,-k\,\cos\theta\,\right)$\\[1ex]
 \hline
\multicolumn{6}{|c|}{} \\*[-2ex]
\multicolumn{6}{|c|}{Centre-of-Mass Normal spin vectors}\\[0.5ex]\hline
& & & & & \\[-1ex]
$ S_1^{\,N} $&$  =$&$ \left(\,0,\,0,\,1,\,0\,\right)$& $ S_3^{\,N} $&$ =$ &$ \left(\,0,\,0\,,1\,,\,0\,\right)$\\[1ex]
$ S_2^{\,N} $&$ = $&$\left(\,0,\,0,\,1,\,0\,\right)$&$ S_4^{\,N} $&$ = $&$ \left(\,0,\,0,\,1,\,0\,\right)$\\[1ex]
 \hline
\multicolumn{6}{|c|}{} \\*[-2ex]
\multicolumn{6}{|c|}{Centre-of-Mass Transverse spin vectors}\\[0.5ex]\hline
& & & & & \\[-1ex]
$  S_1^{\,T} $&$  =$&$ \,\left(\,0,\,1,\,0,\,0\,\right)$& $ S_3^{\,T} $&$ =$ &$ \left(\,0,\,\cos\theta,\,0,\,-\sin\theta\,\right)$\\[1ex]
$ S_2^{\,T} $&$ = $&$ \left(\,0,\,1,\,0,\,0\,\right)$&$ S_4^{\,T} $&$ = $&$ \left(\,0,\,-\cos\theta,\,0,\,\sin\theta\,\right)$\\[1ex]
\hline
\multicolumn{6}{|c|}{} \\*[-2ex]
\multicolumn{6}{|c|}{Centre-of-Mass Longitudinal spin vectors} \\[0.5ex]\hline
& & & & & \\[-1ex]
$  S_1^{\,L} $&$  =$&$ \displaystyle{\frac{1}{M}}\,\left(\,k,\,0,\,0,\,E_1\,\right)$& $ S_3^{\,L} $&$ =$ &$ \displaystyle{\frac{1}{M}}\,\left(\,k,\,E_1\,\sin\theta,\,0,\,E_1\,\cos\theta\,\right)$\\[2ex]
$ S_2^{\,L} $&$ = $&$ \displaystyle{\frac{1}{m}}\,\left(\,k,\,0,\,0,\,-E_2\,\right)$&$ S_4^{\,L} $&$ = $&$ \displaystyle{\frac{1}{m}}\,\left(\,k,\,-E_2\,\sin\theta,\,0,\,-E_2\,\cos\theta\,\right)$\\[2ex]
 \hline
\end{tabular}
\end{center}
\end{table}
\vspace*{-3ex}
Here the Centre-of-Mass energies are $E_1 = \displaystyle{\sqrt{k^{\,2} + M^{\,2}}}$ and $E_2 = \displaystyle{\sqrt{k^{\,2} + m^{\,2}}}$.
Define electromagnetic form factors $F_1(q^2)$ and $F_2(q^2)$, with normalization $F_1(0)=1$ and $F_2(0)=\mu - 1$, the anomalous magnetic moment, where $q^2 = t$.  Using Sach's magnetic form factor $G_M = F_1 + F_2$  and Gordon decomposition, the antiproton current is
\begin{equation}
 J^\mu = \,e_{\bar p} \ \bar u(p_3,\Lambda') \left( G_M \, \gamma^{\,\mu} 
- F_2\,\, \displaystyle{\frac{p_2^{\,\mu} + p_4^{\,\mu}}{2\,M}} \,\right)u(p_1,\Lambda) 
\end{equation}
where $M$ is the mass of the antiproton.
For the spin vectors $S_i$ we use the normalization $S^{\,\mu} S_{\,\mu} = -1$ and $S^{\,\mu} p_{\,\mu} = 0$.  The metric $g^{\,\mu \nu}= \mbox{diag}(\,1,-1,-1,-1\,)$ is used, $\slashed{S} = \gamma^\mu S_\mu$ and $\alpha = e^2/(4\,\pi)$.

\pagebreak

\section{The generic calculation}
\noindent
The generic equation for polarization effects in elastic spin 1/2 - spin 1/2 scattering to first order in QED is
\vspace*{-3ex}
\begin{eqnarray}
&  16\ \displaystyle{\left(\ \frac{q}{e}\ \right)^4}\, |\,\mathcal{M}\,|^{\,2} = \\[1ex]&
\mbox{Tr}\left[\left(\slashed{p}_4 + m \right)\left(1
  +
  \gamma_5 \, \slashed{S}_4 \right) \left(g_M \gamma^\nu + f r^\nu \right) \left(\, \slashed{p}_2
  +
  m\right) \left(1+\gamma_5 \,  \slashed{S}_2 \right) \left(g_M \gamma^\mu + f r^\mu \right)\right] \times \nonumber \\[1ex] 
& 
  \mbox{Tr}\left[\left(\,\slashed{p}_1 + M \right)\left(1
  +
  \gamma_5 \,\slashed{S}_1 \right) \left(G_M \gamma_\mu
  +
  F R_\mu \right)\left(\, \slashed{p}_3 + M \right)\left(1
  +
  \gamma_5 \,  \slashed{S}_3\right)\left(G_M \gamma_\nu + F R_\nu \right)
  \right] \nonumber
\end{eqnarray}
where $\displaystyle{R^{\,\mu} = \left(\,p_1 + p_3\,\right)^{\,\mu}}$, $\displaystyle{r^{\,\mu} = \left(\,p_2 + p_4\,\right)^{\,\mu}}$, $\displaystyle{g_M = f_1 + f_2}$, $\displaystyle{F = -\,F_2 \,/\, (2\,M)}$ and $\displaystyle{f = -\,f_2\, /\, (2\,m)}$.
This generic equation can thus be used to calculate all helicity amplitudes and observables etc. by substituting specific values for the spin ($S_i$) and momenta ($p_i$) four vectors, and can describe equal particle scattering in the case $f_1 \rightarrow F_1$, $f_2 \rightarrow F_2$ and $m \rightarrow M$.  This also applies to electron-proton scattering by setting $f_1 \rightarrow 1$ and $f_2 \rightarrow 0$ and  hence $g_M \rightarrow 1$.  In this case the current in equation (3) would be the usual electron current $j^{\,\mu}= e\,\bar{u}\left(\,p_4,\,\lambda'\,\right)\,\gamma^{\,\mu}\,u\left(\,p_2,\,\lambda\,\right)$.  

Equation (4) can be generalized to directly evaluate the squares of all electromagnetic helicity amplitudes.  We introduce the constants $\epsilon_i$ multiplying each $S_i$, where $i \in \{1,2,3,4\}$.
\begin{eqnarray}
&  16\ \displaystyle{\left(\ \frac{q}{e}\ \right)^4\, |\,\mathcal{M}\,|^{\,2}} = \\[1ex]&
\mbox{Tr}\left[\left( \slashed{p}_4\! + m \right)\left(1\!
  +
  \epsilon_4\,\gamma_5 \, \slashed{S}_4 \right) \left(g_M \gamma^\nu + f r^\nu \right) \left( \slashed{p}_2\!
  +
  m\right) \left(1+\epsilon_2\,\gamma_5 \, \slashed{S}_2 \right) \left(g_M \gamma^\mu + \!f r^\mu \right)\right] \!\times \nonumber \\[1ex] 
& 
\hspace*{-0.12cm}
  \mbox{Tr}\left[\left( \slashed{p}_1\! + \!M \right)\left(1\!
  +
  \epsilon_1\,\gamma_5 \, \slashed{S}_1 \right) \left(G_M \gamma_\mu
  +\!
  F R_\mu \right)\left( \slashed{p}_3 \!+ M \right)\left(1
  \!+
  \epsilon_3\,\gamma_5 \, \slashed{S}_3\right)\!\left(G_M \gamma_\nu +\! F R_\nu \right)
  \right] \nonumber
\end{eqnarray}
Later we will set $\epsilon_i = \pm 1$ to account for different helicity states.  Equation (5) is used to derive the helicity amplitudes and spin observables. 
The spin four vectors are now normalized so that all $\epsilon_i= +1$ corresponds to the helicity amplitude $\phi_1 = \mathcal{M}(\,+,+\,;\,+,+\,)$, and the $\pm 1$ in the helicity amplitudes now relate to the signs of the $\epsilon_i$.

\section{Spin averaged differential cross-section}
\noindent
The spin averaged equation from reference \cite{O'Brien:2006tp} can be simplified using the Sachs electric form factors $\displaystyle{G_E = F_1 + \frac{t}{4\,M^{\,2}}\,F_2}$ and $\displaystyle{g_E = f_1 + \frac{t}{4\,m^2}\,f_2}$ giving:
\begin{eqnarray}
\frac{s}{\alpha^{\,2}}\,\frac{d\,\sigma}{d\,\Omega}
& =
&
\left(\frac{4\,m^2\,g_E^{\,2} -t\,g_M^{\,2}}{4\,m^2 - t}\right)\left(\frac{4\,M^{\,2}\,G_E^{\,2} -t\,G_M^{\,2}}{4\,M^{\,2} - t}\right)\frac{\left(\,M^{\,2}-m^2\,\right)^2 - s\,u}{t^2} \nonumber\\[1ex]
& & +\, \left(\frac{2\,m\,Mg_E\,G_E}{t}\right)^2 + \frac{1}{2}\,\,g_M^{\,2}\,G_M^{\,2}\,.
\end{eqnarray}
\section{Electromagnetic helicity amplitudes}
\noindent
For elastic spin 1/2 - spin 1/2 scattering of non-identical particles there are six independent electromagnetic helicity amplitudes \cite{Buttimore:1978ry,LaFrance:1980}, these are
\begin{center}
\begin{tabular}{ccccccc}
$\phi_1$ & $\equiv$ & $\mathcal{M}(+,+\,;+,+)$ & \ \ \ \ & $\phi_2$ & $\equiv$ & $\mathcal{M}(+,+\,;-,-)$  \\[1ex]
 $\phi_3$ & $\equiv$ & $\mathcal{M}(+,-\,;+,-)$ & \ \ \ \ & $\phi_4$ & $\equiv$ & $\mathcal{M}(+,-\,;-,+)$ \\[1ex]
 $\phi_5$ & $\equiv$ & $\mathcal{M}(+,+\,;+,-)$ & \ \ \ \ & $\phi_6$ &  $\equiv$ & $\mathcal{M}(+,+\,;-,+)$  
\end{tabular}
\end{center}
Note that for identical particle elastic scattering ($p\,p$ and $\bar{p}\,\bar{p}$ in our case) $\phi_6 = - \phi_5$ by isospin invariance.  Using the generic equation (5) the helicity amplitudes $\phi_1,\,\ldots\,,\phi_6$ can be obtained by substituting in the four longitudinal spin vectors (four pure helicity states) from Table~1 and setting specific values ($\pm 1$) for the $\epsilon$'s.  As in references \cite{Buttimore:1978ry,Leader:2005}, we use the notation  $\mathcal{M}\left(\,\mbox{scattered},\,\mbox{recoil}\,;\,\mbox{beam},\,\mbox{target}\,\right)$  i.e. $\mathcal{M}\left(\,\epsilon_3,\,\epsilon_4\,;\,\epsilon_1,\,\epsilon_2\,\right)$.

The electromagnetic helicity amplitudes for the one photon exchange elastic scattering of two spin 1/2 particles with form factors are found to be 
\begin{eqnarray}
\frac{\phi_1}{\alpha} & = & \frac{s - m^2 - M^{\,2}}{t}\left(1\!+\!\frac{t}{4\,k^2}\right) f_1\,F_1 - f_1\,F_1 - f_2\,F_1 - f_1\,F_2 - \frac{1}{2}\,f_2\,F_2\left(\!1 - \frac{t}{4\,k^2}\right) \nonumber \\[2ex]
\frac{\phi_2}{\alpha} & = & \! \frac{1}{2}\left(\frac{m}{k}\,f_1 - \frac{k}{m}\,f_2\right)\!\left(\frac{M}{k}\,F_1 - \frac{k}{M}\,F_2\right)\!+ \frac{s - m^2\! - M^{\,2}\! - 2\,k^2}{4\,m\,M}\left(\!1+\!\frac{t}{4\,k^2}\right) f_2\,F_2\nonumber \\[2ex]
\frac{\phi_3}{\alpha} & = & \left(\frac{s - m^2 - M^{\,2}}{t}\, f_1\,F_1 + \frac{f_2\,F_2}{2}\,\right)\left(1 + \frac{t}{4\,k^2}\right)  \\[2ex]
\phi_4 & = & -\phi_2 \nonumber \\[2ex]
\frac{\phi_5}{\alpha} & = &\! \sqrt{\frac{s\left(4\,k^2 + t\right)}{-t}}\left[\frac{f_1\,F_1\,m}{4\,k^2}\left(\!1 \!- \frac{m^2\! - M^{\,2}}{s}\right) \!-\! \frac{F_1\,f_2}{2\,m} \!+\frac{t\,f_2\,F_2}{16\,m\,k^2}\left(\!1 \!+ \frac{m^2\! - M^{\,2}}{s}\right)\!\right]\nonumber \\[2ex]
\frac{\phi_6}{\alpha} & = &\! \sqrt{\frac{s\left(4\,k^2 + t\right)}{-t}}\!\left[\frac{f_1\,F_1\,M}{4\,k^2}\!\left(\frac{\!M^{\,2}\! - m^2}{s}-1\!\right) \!+\!\frac{F_2\,f_1}{2\,M} -\frac{t\,f_2\,F_2}{16\,M\,k^2}\left(\!1 \!+ \frac{M^{\,2}\! - m^2}{s}\right)\!\right]\nonumber 
\end{eqnarray}
in agreement with those found by other methods \cite{Buttimore:1978ry,LaFrance:1980}.
 The Centre-of-Mass three momentum $k$ obeys $4\,k^2\,s = [\,s - \left(M+m\right)^2\,]\,[\,s- \left(M-m\right)^2\,]$.\\
The combinations $\left(\,\phi_1 + \phi_3\,\right)$ and $\left(\,\phi_1 - \phi_3\,\right)$ appear often in the observables
\begin{eqnarray}
\frac{\phi_1 + \phi_3}{\alpha} & = & -\,g_M\,G_M + \left(\,1+\frac{t}{4\,k^2}\right)\,\left(\frac{s - m^2 - M^{\,2}}{t}\,2\,f_1\,F_1 + \,f_2\,F_2\,\right)\\[2ex]
\frac{\phi_1 - \phi_3}{\alpha} & = & -\,g_M\,G_M\,.
\end{eqnarray}

\section{Polarization transfer observables}
\noindent
Setting $S_1 = S_4 =0$ in equation (5) gives a generic spin transfer equation, into which specific vectors $S_2$ and $S_3$ will be inserted to give the various polarization transfer observables.  Scattering is in the $XZ$ plane, so coordinates are $X$ (Transverse), $Y$ (Normal) and $Z$ (Longitudinal), which are related to  the Argonne LAB coordinates $S$, $N$ and $L$ \cite{Leader:2005}.  
\subsection{The spin observables $K_\mathrm{XX}$, $K_\mathrm{YY}$ and $K_\mathrm{ZZ}$}
\noindent
The spin observable $K_\mathrm{XX}$ is obtained by inserting the transverse polarized spin vectors $S_2 = S_2^{\,T}$ and $S_3 = S_3^{\,T}$ from Table~1 into the generic spin transfer equation:
\begin{eqnarray}
\frac{d\,\sigma}{d\,\Omega}\,K_\mathrm{XX} & = & \alpha^2\frac{g_M\,G_M}{8\,k^2\,m\,M\,s}\,\left\{\,4\,m^2\,f_1\,\left(\,M^{\,2}\,F_1 - k^2\,F_2\,\right) + f_2\,\left[-4\,k^2\,M^{\,2}\,F_1 \right. \right. \nonumber \\[1ex]
 & &  \left. \left. \qquad +\,\left(\,4\, k^4 + \left(\,4\,k^2\,+ t\,\right)\sqrt{k^2 + m^2}\,\sqrt{k^2 + M^{\,2}}\,\right)\, F_2\,\right]\right\}\,.
\end{eqnarray}
Inserting the normal polarized spin vectors $S_2 = S_2^{\,N}$ and $S_3 = S_3^{\,N}$ from Table~1 into the generic spin transfer equation gives
\begin{equation}
\frac{d\,\sigma}{d\,\Omega}\,K_\mathrm{YY} = \left(\frac{2\,\alpha^2}{s\,t}\right)\,\,m\,M \,g_E\,g_M\,G_E\,G_M \,.
\end{equation}
Inserting the longitudinally polarized spin vectors $S_2 = S_2^{\,L}$ and $S_3 = S_3^{\,L}$ from Table~1 into the generic spin transfer equation gives
\begin{eqnarray}
\frac{d\,\sigma}{d\,\Omega}\,K_\mathrm{ZZ} & = & -\,\alpha^2\,\frac{g_M\,G_M\,}{8\ k^2\,s^2\, t}\left\{\left[s^2 - \left(M^{\,2} - m^2\right)^2\right]\,\left(4\,k^2
+ t\right)\, f_1\, F_1 \right. \nonumber\\[1ex]
& &  \left. 
\phantom{\sqrt{k^2 + M^{\,2}}} \qquad \qquad
+\,s\,\left(4\,k^2\,f_1 - t\, f_2\right)\,\left(\,4\,k^2\, F_1 - t\,F_2\,\right)\right\}\,.
\end{eqnarray}
\vspace*{-4ex}
\subsection{The spin observables $K_\mathrm{XZ}$ and $K_\mathrm{ZX}$}
\noindent
When the spin four vectors $S_2 = S_2^{\,T}$ and $S_3 = S_3^{\,L}$ from Table~1 are substituted into the generic spin transfer equation we obtain
\begin{eqnarray}
\frac{d\,\sigma}{d\,\Omega}\,K_\mathrm{XZ} & = & \frac{\alpha^2\,g_M\,G_M}{16\,m\,\,s^{\,3/2}\,t}\sqrt{\frac{-\,t\,\left(\,4\,k^2 + t\,\right)}{k^{\,4}}}\,\left\{\,\left(\,s + m^2 - M^{\,2}\,\right)\,t\,f_2\,F_2 \right. \nonumber \\[1ex]
& & \left. \qquad \qquad \qquad +\ 4\,F_1\,\left[\,m^2\,\left(s + M^2 - m^2\right)\,f_1 - 2\,k^2\,s\,f_2\,\right] \right\}\,.
\end{eqnarray}
Inserting the spin four vectors $S_2 = S_2^{\,L}$ and $S_3 = S_3^{\,T}$ from Table~1  we obtain
\begin{eqnarray}
\frac{d\,\sigma}{d\,\Omega}\,K_\mathrm{ZX} & = &  \frac{\alpha^2\,g_M\,G_M}{16\,M\,s^{\,3/2}\,t}\sqrt{\frac{-\,t\,\left(\,4\,k^2 + t\,\right)}{k^{\,4}}}\,\left\{\left(\,s + M^{\,2} - m^2\,\right)\,t\,f_2\,F_2 \right.\nonumber \\[1ex]
& & 
\left. \qquad \qquad \qquad +\  4\,f_1\,\left[\,M^2\,\left(s + m^2 - M^2\,\right)F_1 - 2\,k^2\,s\,F_2\,\right]\right\}\,.
\end{eqnarray}
As expected by parity and time reversal invariance we confirm that
\begin{equation}
K_\mathrm{XY} \ = \ K_\mathrm{YX} \ = \ K_\mathrm{YZ} \ = \ K_\mathrm{ZY} \ = \ 0\,.
\end{equation}
\section{Depolarization spin observables}
\noindent
The observable $D_{ij}$ is the polarization remaining after interaction with the target, of interest here is the loss of polarization after interaction with the target, i.e. $\left(1 - D_{ij}\right)$.  We present results to leading order in small $t$.  Here setting $S_2 = S_4 = 0$ in the generic equation (5) and subtracting from the spin averaged equation gives a generic depolarization equation, into which the various vectors $S_1$ and $S_3$ can be substituted.  The $\approx$ sign means to first order in small $t$.
\subsection{The spin observables $\left(\,1 - D_\mathrm{XX}\,\right)$, $\left(\,1 - D_\mathrm{YY}\,\right)$ and $\left(\,1 - D_\mathrm{ZZ}\,\right)$}
\noindent
Substituting the transverse spin vectors $S_1 = S_1^{\,T}$ and $S_3 = S_3^{\,T}$ from Table~1 into the generic depolarization equation gives
\begin{eqnarray}
\frac{d\,\sigma}{d\,\Omega}\,\left(\,1 - D_\mathrm{XX}\,\right) & \approx & \frac{-\,2\,\alpha^2\,F_1^2}{k^2\, m^2\,t\,s}\, \left\{ \left[\,m^4\ \left(\,k^2 + M^{\,2}\,\right)\, f_1^2  + s\,k^{\,4}\,f_2^{\,2} \right.\right.\\[1ex] 
& & \qquad \qquad \qquad \left. \left.-\ k^2\ m^2\ \left(s + M^{\,2} - m^2\,\right)\, f_1\ f_2 \nonumber
\right]\right\}\,.
\end{eqnarray}
Inserting the normal polarized spin vectors $S_1 = S_1^{\,N}$ and $S_3 = S_3^{\,N}$ from Table~1 into the generic depolarization equation gives
\begin{equation}
\frac{d\,\sigma}{d\,\Omega}\,\left(\,1 - D_\mathrm{YY}\,\right)  = \frac{\alpha^2}{2\,s}\, g_M^{\,2}\ G_M^{\,2}\,, \ \ \ \ \ \ \ \ \mbox{complete to all orders in $t$.}
\end{equation}
Inserting the longitudinal spin vectors $S_1 = S_1^{\,L}$ and $S_3 = S_3^{\,L}$ from Table~1 into the generic depolarization equation gives
\begin{eqnarray}
\frac{d\,\sigma}{d\,\Omega}\,\left(\,1 - D_\mathrm{ZZ}\,\right) & \approx & \frac{-\,2\,\alpha^2\,F_1^2}{k^2\, M^2\,t\,s}\, \left\{ \left[\,M^4\ \left(k^2 + m^2\right)\, f_1^2 + s\,k^{\,4}\,f_2^{\,2} \right.\right.\\[1ex] 
& & \qquad \qquad \qquad \left. \left.-\ k^2\ M^2\ \left(s+m^2 - M^{\,2}\,\right)\, f_1\ f_2 \nonumber
\right]\right\}\,.
\end{eqnarray}
\subsection{The spin observables $\left(\,1 - D_\mathrm{XZ}\,\right)$ and $\left(\,1 - D_\mathrm{ZX}\,\right)$}
\noindent
Substituting the spin vectors $S_1 = S_1^{\,T}$ and $S_3 = S_3^{\,L}$ from Table~1 into the generic depolarization equation gives
\begin{equation}
\frac{d\,\sigma}{d\,\Omega}\,\left(\,1 - D_\mathrm{XZ}\,\right)   \approx  \frac{4\ \alpha^2\ f_1^{\,2}\ F_1^{\,2}}{s\ t^{\,2}} \left[\ 2\ k^{\,4} + m^2\ M^{\,2}  + \  k^{\,2}\ \left(\,s - 2\,k^{\,2}\,\right)\,\right]
\end{equation}
and $\displaystyle{\frac{d\,\sigma}{d\,\Omega}\,\left(\,1 - D_\mathrm{ZX}\,\right)}$, found by  inserting into the generic depolarization equation the spin vectors $S_1 = S_1^{\,L}$ and $S_3 = S_3^{\,T}$ from Table~1, is the same as above to leading order in $t$.  Note this is just the leading $t$ part of the spin averaged case.

As expected by parity and time reversal invariance we confirm that
\begin{equation}
D_\mathrm{XY} \ = \ D_\mathrm{YX} \ = \ D_\mathrm{YZ} \ = \ D_\mathrm{ZY} \ = \ 0\,.
\end{equation}

\section{Spin asymmetries}
\noindent
For electromagnetic scattering to first order in QED all single and triple spin asymmetries are zero,
\begin{equation}
A_{i} = A_{ijk} = 0\,\,\,\,\,\mbox{where $i$, $j$, $k$}\, \in \{\mathrm{\,X,Y,Z\,}\} 
\end{equation}
and all the double spin asymmetries equal the polarization transfer spin observables.
\begin{equation}
A_{ij} = K_{ij}\,\,\,\,\,\mbox{where $i$, $j$}\, \in \{\mathrm{\,X,Y,Z\,}\} \,.
\end{equation}

\section{Four spin measurement spin observables}
\noindent
There are three independent four spin measurement spin observables $\left(\,\mathrm{XX}\,|\,\mathrm{XX}\,\right)$, $\left(\,\mathrm{XX}\,|\,\mathrm{XZ}\,\right)$ and $\left(\,\mathrm{XX}\,|\,\mathrm{ZX}\,\right)$ \cite{Leader:2005}.  
All other four spin measurement spin observables are related to the spin observables presented in earlier sections \cite{LaFrance:1980}.  These are not needed for the $\mathcal{PAX}$ collaboration as the polarization of the recoil particle will not be measured.

\section{Spin observables for  $p\,p$, $\bar{p}\,p$ and $\bar{p}\,\bar{p}$ scattering}
\noindent
The electromagnetic helicity amplitudes and spin observables for $p\,p$, $\bar{p}\,p$ and $\bar{p}\,\bar{p}$ scattering can be obtained by setting equal masses and form factors ($f_1 = F_1$, $f_2 = F_2$, $g_E = G_E$, $g_M = G_M$ and $m = M$) in the expressions provided in previous sections.  These are required by $\mathcal{PAX}$ to analyze the buildup of polarization of an antiproton beam by interactions with the protons in a hydrogen target.  The helicity amplitudes in section 5 now become
\begin{eqnarray}
\frac{\phi_1}{\pm \,\alpha} & = & \left(\frac{s + 4\,k^2}{2\,t} + \frac{M^{\,2}}{2\,k^2}\right)\,F_1^{\,2} - 2\,F_1\,F_2 + \left(\frac{t - 4\,k^2}{8\,k^2}\right)\,F_2^{\,2} \nonumber \\[2ex]
\frac{\phi_2}{\pm \,\alpha} & = &  \frac{-\,\phi_4}{\pm \,\alpha}  =  \left(\frac{M^{\,2}}{2\,k^2}\right)\,F_1^{\,2} - F_1\,F_2 + \left[\frac{s\left(\ t+8\,k^2\ \right)}{32\,M^{\,2}\,k^2}-\frac{1}{2}\right]\,F_2^{\,2} \nonumber \\[2ex]
\frac{\phi_3}{\pm \,\alpha} & = & \left(\frac{s - 2\,M^{\,2}}{t}\,F_1^{\,2}+\frac{F_2^{\,2}}{2} \right)\left(\,1 + \frac{t}{4\,k^2} \right)\\[2ex]
\frac{\phi_5}{\pm \,\alpha} & = & \frac{1}{2\,M}\sqrt{\frac{\,s\,\left(4\,k^2 +t\right)}{-t}}\,\left[\left(\frac{M^{\,2}}{2\,k^2}\right)\,F_1^{\,2} - F_1\,F_2 + \left(\frac{t}{8\,k^2}\right)\,F_2^{\,2}\, \right]\nonumber
\end{eqnarray}
where $\phi_6 = - \phi_5$ in this case.  The factor of $\pm$ difference between here and section 5 is from the opposite charge of the antiproton $e_p = - e_{\bar{p}}$ so $e_p\,e_{\bar{p}} = -e_p^{\,2}$.  So of the $\pm$ signs in this section the $+$ sign relates to $p\,p$ and $\bar{p}\,\bar{p}$ scattering and the $-$ sign relates to $\bar{p}\,p$ scattering.

The spin transfer observables now become
\begin{eqnarray}
\frac{d\,\sigma}{d\,\Omega}\,K_\mathrm{XX} & =\! &\! \frac{\alpha^2\,G_M^{\,2}}{8\,s\,k^{\,2}\,M^2}\!\left\{\!4\,M^{\,4}F_1^{\,2}\!- 8\,k^2 M^{\,2}F_1\,F_2 
+\!\left[4\, k^4 \!+ \!\left(\!k^2\!+ \!\frac{t}{4}\right) s\right]\! F_2^{\,2}\right\}\nonumber\\[2ex]
\frac{d\,\sigma}{d\,\Omega}\,K_\mathrm{YY} & = & \left(\frac{2\,\alpha^2}{s\,t}\right)\,M^{\,2} \,G_E^{\,2}\,G_M^{\,2} \\[2ex]
\frac{d\,\sigma}{d\,\Omega}\,K_{ZZ} & = & \frac{-\,\alpha^2\,G_M^{\,2}}{8\ k^2\,s\,\,t}\left[\,s\left(4\,k^2
+ t\,\right)F_1^{\,2}
+\left(\,4\,k^2\, F_1 - t\,F_2\,\right)^2\,\right] \nonumber\,.
\end{eqnarray}
For $p\,p$, $\bar{p}\,p$ and $\bar{p}\,\bar{p}$ elastic scattering $K_{XZ} = K_{ZX}$ thus we obtain
\begin{equation}
\!\frac{d\,\sigma}{d\,\Omega}\,K_{XZ} \! =\!  \frac{d\,\sigma}{d\,\Omega}\,K_{ZX} \! = \!\frac{\alpha^2\,G_M^{\,2} \sqrt{s}}{2\,M\,t}\,\sqrt{\frac{-\,t\,\left(\,4\,k^2 + t\,\right)}{k^{\,4}}}\left(\frac{M^2\,F_1^{\,2}}{2}  - k^2\,F_1\,F_2 + \frac{t\,F_2^{\,2}}{8} \right)
\end{equation}
The depolarization spin observables to leading order in small $t$ for this case are
\begin{eqnarray}
\frac{d\,\sigma}{d\,\Omega}\,\left(\,1 - D_\mathrm{XX}\,\right) & \approx &  \frac{-\,2\,\alpha^2\,F_1^{\,2}}{k^{\,2}\,M^{\,2}\,s\,t}\,\left(\,k^{\,2} + M^{\,2}\,\right)\left(\,M^{\,2}\,F_1 - 2\,k^2\,F_2\,\right)^2 \nonumber\\[2ex]
\frac{d\,\sigma}{d\,\Omega}\,\left(\,1 - D_\mathrm{YY}\,\right) & = & \frac{\alpha^2}{2\,\,s} \, G_M^{\,4}\,,  \ \ \ \ \ \ \ \ \mbox{complete to all orders in $t$} \nonumber\\[2ex]
\frac{d\,\sigma}{d\,\Omega}\,\left(\,1 - D_\mathrm{ZZ}\,\right) & \approx & \frac{-\,2\,\alpha^2\,F_1^{\,2}}{k^{\,2}\,M^{\,2}\,s\,t}\,\left(\,k^{\,2} + M^{\,2}\,\right)\left(\,M^{\,2}\,F_1 - 2\,k^2\,F_2\,\right)^2\\[2ex]
\frac{d\,\sigma}{d\,\Omega}\,\left(\,1 - D_\mathrm{XZ}\,\right) & \approx & \frac{d\,\sigma}{d\,\Omega}\,\left(\,1 - D_\mathrm{ZX}\,\right)  \approx \frac{d\,\sigma}{d\,\Omega} \approx \frac{4\ \alpha^2\ F_1^{\,4}}{s\ t^{\,2}}\left(\,2\ k^{\,2} + M^{\,2}\,\right)^{\,2} \nonumber\,.
\end{eqnarray}
Note that to first order in small $t$ the (anti)proton form factors can be approximated as $F_1 \approx 1$, $F_2 \approx \mu - 1$ (i.e. $G_M \approx \mu$ and $G_E \approx 1$), where $\mu$ is the magnetic moment of the (anti)proton.

The spin averaged differential cross-section (eq. (6)) for this case simplifies to
\begin{equation}
\frac{s}{\alpha^2}\,\frac{d\,\sigma}{d\,\Omega} =
\left(\frac{4\,M^{\,2}\,G_E^{\,2} -t\,G_M^{\,2}}{4\,M^{\,2} - t}\right)^2 \left(\frac{- s\,u}{t^2}\right) + \left(\frac{2\,M^{\,2}\,G_E^{\,2}}{t}\right)^2 + \frac{1}{2}\,\,G_M^{\,4}\,.
\end{equation}
\section{Spin observables for $p\,e$ and $\bar{p}\,e$ scattering}
\noindent
The spin observables for $p\,e$ and  $\bar{p}\,e$ scattering can be obtained by setting $f_1 = 1$ and  $f_2 = 0$ (i.e $g_E = g_M = 1$) in the expressions from sections 4 to 8.  These are required by $\mathcal{PAX}$ to analyze the buildup of polarization of an antiproton beam by interactions with the electrons in a hydrogen target.  Of the $\pm$ signs below the $+$ signs relate to $\bar{p}\,e$  scattering and the $-$ signs relate to $p\,e$ scattering.
\pagebreak

The helicity amplitudes in section 5 for baryon electron collisions become
\begin{eqnarray}
\frac{\phi_1}{\pm \,\alpha} & = & \left(\,s - m^2 - M^{\,2}\,\right) \left(1 + \frac{t}{4\ k^{\,2}}\right) \frac{F_1}{t} - F_1 - F_2\nonumber \\[2ex]
\frac{\phi_2}{\pm \,\alpha} & = &   \frac{-\,\phi_4}{\pm \,\alpha}  = \frac{m\ M\ F_1}{2\ k^{\,2}} - \frac{m\ F_2}{2\ M}\nonumber \\[2ex]
\frac{\phi_3}{\pm \,\alpha} & = &\left(\,s - m^2 - M^{\,2}\,\right) \left(1 + \frac{t}{4\ k^{\,2}}\right) \frac{F_1}{t} \\[2ex]
\frac{\phi_5}{\pm \,\alpha} & = & \sqrt{\frac{s\ \left(\,4\ k^2 + t\,\right)}{-\,t}}\ \left[\frac{m\ \left(\,s - m^2 + M^{\,2}\,\,\right)
 F_1}{4\ k^{\,2}\ s} \right]\nonumber\\[2ex]
\frac{\phi_6}{\pm \,\alpha} & = & -\,\sqrt{\frac{s\, \left(\,4\ k^2 + t\,\right)}{-\,t}}\ \left[\frac{M\ \left(\,s +m^2 - M^{\,2}\,\right) 
F_1}{4\ k^{\,2}\ s} - \frac{F_2}{2\, M}\right]\nonumber\,.
\end{eqnarray}
The spin transfer observables for baryon electron elastic scattering are
\begin{eqnarray}
\frac{d\,\sigma}{d\,\Omega}\,K_{XX} & = &  \alpha^2\frac{m\ G_M}{2\,k^{\,2}\,M\,s}\,\left(\,M^{\,2}\,F_1 - k^{\,2}\,F_2\,\right)\nonumber \\[2ex]
\frac{d\,\sigma}{d\,\Omega}\,K_{YY} & = &  \left(\frac{2\,\alpha^2}{s\ t}\right)\,\,m\,M \,G_E\,G_M \\[2ex]
\frac{d\,\sigma}{d\,\Omega}\,K_{ZZ} & =\! &\! \frac{-\,\alpha^2\,G_M\,}{8\ k^{\,2}\,s^2\ t}\!\left\{\left[s^2 - \left(M^{\,2} - m^2\right)^2\,\right]\,\left(4\,k^2
+ t\,\right) F_1
+4\,k^2\,s\,\left(4\,k^2 F_1 - t\,F_2\right)\right\}\nonumber \\[2ex]
\frac{d\,\sigma}{d\,\Omega}K_{XZ} & = & \frac{\alpha^2\,m\,F_1\,G_M}{4\,s^{\,3/2}\ t}\,\sqrt{\frac{-\,t\,\left(\,4\,k^2 + t\,\right)}{k^{\,4}}}\,\,\left(\,s  - m^2 + M^{\,2}\,\right) \nonumber \\[2ex]
\frac{d\,\sigma}{d\,\Omega}\,K_{ZX} & = & \frac{\alpha^2\,G_M}{4\,M\,s^{\,3/2}\ t}\sqrt{\frac{-\,t\left(4\,k^2 + t\,\right)}{k^{\,4}}}\left[M^2\left(\,s + m^2\!\! - M^2\right)F_1\! - 2\,k^2s\,F_2\,\right]\nonumber\,.
\end{eqnarray}
The depolarization spin observables to leading order in small $t$ for this case are
\begin{eqnarray}
\frac{d\,\sigma}{d\,\Omega}\,\left(\,1 - D_\mathrm{XX}\,\right) & \approx & \frac{-m^2\,\alpha^2\,F_1^{\,2}}{2\ k^2\ s^2\ t}\ \left(\,s - m^2 + M^{\,2}\,\right)^2 \nonumber\\[2ex]
\frac{d\,\sigma}{d\,\Omega}\,\left(\,1 - D_\mathrm{YY}\,\right) & = & \frac{\alpha^2}{2\,\,s} \, G_M^{\,2}\,,  \ \ \ \ \ \ \ \ \mbox{complete to all orders in $t$} \nonumber\\[2ex]
\frac{d\,\sigma}{d\,\Omega}\,\left(\,1 - D_\mathrm{ZZ}\,\right) & \approx & \frac{-M^2\,\alpha^2\,F_1^{\,2}}{2\ k^2\ s^2\ t}\ \left(\,s + m^2 - M^{\,2}\,\right)^2\\[2ex]
\frac{d\,\sigma}{d\,\Omega}\,\left(\,1 - D_\mathrm{XZ}\,\right) & \approx & \frac{d\,\sigma}{d\,\Omega}\,\left(\,1 - D_\mathrm{ZX}\,\right)  \approx \frac{d\,\sigma}{d\,\Omega} \approx \frac{4\ \alpha^2\  F_1^{\,2}}{s\ t^{\,2}}\ \left(\ s\ k^{\,2} + m^2\ M^{\,2}\,\right)  \nonumber\,.
\end{eqnarray}
The spin averaged differential cross-section from section 4 for this case simplifies to
\begin{equation}
\frac{s}{\alpha^2}\,\frac{d\,\sigma}{d\,\Omega}
 =
\left(\frac{4\,M^{\,2}\,G_E^{\,2} -t\,G_M^{\,2}}{4\,M^{\,2} - t}\right)\frac{\left(\,M^{\,2}-m^2\,\right)^2 - s\,u}{t^2} + \left(\frac{2\,m\,M\,G_E}{t}\right)^2 + \frac{1}{2}\,\,\,G_M^{\,2}
\end{equation}
the familiar Rosenbluth formula.

\section{Summary}
\noindent
All electromagnetic helicity amplitudes and spin observables for elastic antiproton electron and antiproton proton scattering have been presented, the depolarization observables only to leading order in small $t$.  These are important for the $\mathcal{PAX}$ collaboration, which plans to polarize an antiproton beam by repeated interaction with a hydrogen target in a storage ring.  In order for many of the beam particles to remain in the ring after scattering, small momentum transfer $t$ is considered.  The spin observables can be used to estimate the rate of buildup of polarization of an antiproton beam by spin filtering \cite{Nikolaev:2006gw,Milstein:2005bx}.  Spin filtering requires evaluation of the angular integration of the product of the observables $A_{ii} = K_{ii}$ and $\left(1 - D_{ii}\right)$ with $d\,\sigma / d\,\Omega$.  Azimuthal averaging indicates that the observables with single $\mathrm{X}$ (i.e. $K_\mathrm{XZ}$, $K_\mathrm{ZX}$, $D_\mathrm{XZ}$ and $D_\mathrm{ZX}$) do not contribute to spin filtering.  The quantities $\left(K_\mathrm{XX} + K_\mathrm{YY}\right)/2$, $\left(D_\mathrm{XX} + D_\mathrm{YY}\right)/2$, $K_\mathrm{ZZ}$ and $D_\mathrm{ZZ}$ play the important role.   General results for helicity amplitudes and spin observables for any elastic spin 1/2 - spin 1/2 scattering have also been presented to first order in QED.

 \bigskip

 {\small DOB would like to thank \lq\lq The Embark Initiative'' (IRCSET), for a postgraduate research scholarship.  NHB is grateful to Enterprise Ireland for the award of a grant under the International Collaboration Program to facilitate a visit to INFN at the University of Torino, where discussions with M.~Anselmino and M.~Boglione are acknowledged.  We are also grateful to E.~Leader and N.~N.~Nikolaev for helpful comments.}

 \end{document}